\documentclass[pre,twocolumn,showpacs]{revtex4}
\usepackage{graphicx}
\usepackage{epsfig}
\usepackage{bm}
\usepackage{latexsym}

\begin{document}

\title{Critical Level Statistics of the  Fibonacci Model}

\author{Michihiro Naka$^1$, Kazusumi Ino$^1$ and Mahito Kohmoto$^2$}

\affiliation{
$^1$Department of Pure and Applied Sciences, University of Tokyo, 
Komaba 3-8-1, Meguro-ku, Tokyo, 153-8902, Japan\\
$^2$Institute for Solid State Physics, University of Tokyo,
Kashiwanoha 5-1-5, Kashiwa-shi, Chiba, 277-8581, Japan}

\begin{abstract}
\hspace{5mm}
We numerically analyze spectral properties of the  
Fibonacci model which is a one-dimensional quasiperiodic system.
We find that the energy levels of this model have  
the distribution  of the band widths $w$ obeys 
$P_B(w)\sim w^{\alpha}$ $(w\to 0)$ 
and $P_B(w) \sim e^{-\beta w}$ $(w\to\infty)$,  
the gap distribution $P_G(s)\sim s^{-\delta}$ $(s\to 0)$ 
($\alpha,\beta,\delta >0$) .
We also compare  the results with those of multi-scale Cantor sets. 
We find qualitative differences between  
the spectra of the Fibonacci model and the multi-scale Cantor sets.
\end{abstract}

\pacs{05.45.Pq, 05.70.Jk, 71.23.Ft}

\maketitle
\section{Introduction}
\hspace{2.5mm}
Bloch's theorem tells that 
the  free electron eigenstates are always extended in periodic systems. 
It is also well known that the free electron eigenstates are always 
localized in one-dimensional random media 
\cite{LeeRam}.
On the other hand, one-dimensional quasiperiodic systems 
show various interesting behavior \cite{HirKoh}. 
A one-dimensional tight-binding model is    
\begin{equation}
t_{i+1}\psi_{i+1}+t_{i-1}\psi_{i-1}+\epsilon_i \psi_i =E\psi_i,
\label{2}
\end{equation} 
where $\psi_i $ denotes  the value of the 
wave function at $i$-th site, 
$t_i$ and $\epsilon_i$  are  the hopping matrix element 
and  the site energy at $i$-th site respectively, both of which can   
be taken to be quasiperiodic with lattice spacing. 
The Harper model:
$t_{i}=1$ and $\epsilon_i=\lambda \cos(2\pi\sigma i+\theta)$ has been 
well studied.  
When $\sigma$ is an irrational number, it is quasiperiodic.  
The eigenstates of the quasiperiodic Harper model
 are extended for $\lambda <2$ 
and are localized for $\lambda >2$ with the metal-insulator 
transition point $\lambda=2$ \cite{AubAnd}.  
Hofstadter found a surprisingly rich structure of 
the spectrum at $\lambda = 2$ \cite{Hof}.
There the spectrum as well as the eigenstates become multifractal
\cite{HirKoh}. 
The total measure of the bands at the critical point 
is zero with a fractal dimension less than one \cite{Thouless,Koh}. 
The scaling behavior of the spectrum has been extensively studied   
\cite{HirKoh}
by multifractal analysis \cite{Halsey}. 

Recently a new statistical characterization of the spectrum  
of the Harper model at the critical point was found 
\cite{EvaPic,TakInoYam}. 
The distribution of the normalized width of the bands was examined.
The distribution of the band-widths
$P_B(w)dw$ counts the number 
of bands whose values are between $w$ and $w+dw$.
In \cite{EvaPic},  $P_B(w)$ was numerically investigated 
 at $\lambda=2$ and $\sigma=\frac{\sqrt{5}-1}{2}$ 
and it was  claimed that $P_B(w)$ follows 
the semi-Poissonian distribution $4w e^{-2w}$.  
However this claim has been excluded 
by further detailed numerical investigations \cite{TakInoYam}.
Still it has been confirmed that 
\begin{eqnarray}
P_B(w)\sim w^{\alpha}
\quad (w\to 0),
\end{eqnarray}
and
\begin{eqnarray}
P_B(w)\sim e^{-\beta w}
\quad (w \to \infty),
\end{eqnarray}
where $\alpha \sim 2.5,\beta \sim 1.4$ \cite{TakInoYam}. 
Similar laws are also found for  
a variant of the Harper model at its criticality 
\cite{TakInoYam}.

For the Harper model, 
the distribution of energy gaps $P_G(s)$ was also examined.
The distribution diverges near the origin and 
follows an inverse power law 
\cite{TakInoYam,MachiGei}
\begin{eqnarray}
P_G(s)\sim s^{-\delta} \quad (s\to 0),
\end{eqnarray}
with  $\delta \sim 1.5$.  
This law is also extended to a variant of the Harper 
model \cite{TakInoYam}. 

In order to investigate how these laws are universal for 
critical quasiperiodic systems, we   study  
the one-dimensional Fibonacci model \cite{KohKadTan,Ostlund}. 
It  is  a model obtained by setting $t_i$ or 
$\epsilon_i$ in (\ref{2}) to the Fibonacci sequence. We 
shall study  the case with $\epsilon_i=0$. 
The Fibonacci model  remains critical when varying the parameters $t_i$, 
and the spectra are multifractal similar to that at the critical 
point $\lambda=2$ of the Harper model \cite{HirKoh}.
We will find 
 qualitatively similar but quantitatively different 
behavior of $P_B(w)$ and $P_G(s)$. 

A remarkable feature of this model is that 
the construction of the Fibonacci sequence can be 
translated to  
a renormalization group transformation, and 
its action on the trace of the 
transfer matrix becomes a nonlinear dynamical system called 
the Kohmoto-Kadanoff-Tang (KKT) map\cite{KohKadTan}.   
The KKT map  has been analyzed as a nonlinear dynamical system 
\cite{KohOon,KohBan,KohSutTan}. 
Namely the periodic orbit corresponding to the bands' center was analyzed,
and the scaling property of the band was shown to be determined by 
the eigenvalues of the linearized map at the hyperbolic points. 
They induce heteroclinic points of the KKT map. 
 The structure of subsets of the spectrum was 
explained by the Smale's horseshoe structure.  
This analysis can be extended to other bands and 
then the explanation is generalized to a subset of the whole spectrum 
\cite{KohOon}. 

The Smale's horseshoe structure indicates that 
the spectrum of the Fibonacci model is a multi-scale Cantor set.
Here, we restrict the meaning of the term  
``multi-scale Cantor set'' as the set which is obtained 
by the following procedure : we transform the interval $I=[0,1]$ into 
a finite number of subintervals with definite scales   
and apply the transformation to subintervals successively.  
In the literature, the term ``Cantor set''  is also  
used as the term for  ``a closed set with 
no isolated points and whose complement is dense''.  
We call this  "general Cantor set".  
We will compare our results 
with those of  multi-scale Cantor sets. 
Also note that the spectrum of the Fibonacci model is 
 proved to be
singular continuous \cite{Suto}. 
Nevertheless the comparison will turn out to be useful.

In Sec.II, 
we recall 
the renormalization group transformation.  
In Sec.III, we report the results of our numerical studies. 
In Sec.IV, we compare our results with those of  multi-scale Cantor sets. 
Sec.V gives conclusion.  
In Appendix, we provide definition and distributions of
  multi-scale Cantor sets.

\section{Fibonacci Model and the Kohmoto-Kadanoff-Tang map}
\hspace{2.5mm}
We recall the aspects of the Fibonacci model 
we  use,
especially the renormalization group transformation  \cite{KohKadTan}. 

The Fibonacci model describes
a one-dimensional quasiperiodic system given by (\ref{2}) 
with $\epsilon_i=0$,   defined by 
\begin{eqnarray}
\label{off-diag}
t_{i+1}\psi_{i+1}+t_i\psi_{i-1}=E\psi_i,
\end{eqnarray}
where $t_i$'s take two values $t_a$ and $t_b$ 
arranged in the Fibonacci sequence  
constructed recursively as 
$S_{\ell+1}=\{S_{\ell},S_{\ell-1}\}$ $(\ell\geq 1)$ with 
an initial condition $S_{0}=\{b\}, S_1=\{a\}$.
The number of elements in $S_{\ell}$
is the Fibonacci number $F_{\ell}$ defined by
$F_{\ell+1}=F_{\ell}+F_{\ell-1}$ $(\ell\geq 1)$ 
with an initial condition $F_0=F_1=1$.
To investigate the model with infinite sites,
one considers a finite size system of (\ref{off-diag}) 
with $F_{\ell}$ sites and imposes a boundary condition on the states. 
Then the system is equivalent to the system with a period $F_{\ell}$. 

We write Eq.(\ref{off-diag}) as
\begin{eqnarray}
\left(\begin{array}{c}
\psi_{i+1}\\
\psi_i
\end{array}
\right)
=
\left(
\begin{array}{cc}
\frac{E}{t_{i+1}} & -\frac{t_i}{t_{i+1}}\\
1 &0
\end{array}
\right)
\left(
\begin{array}{c}
\psi_i\\
\psi_{i-1}
\end{array}
\right).
\end{eqnarray}
The present $2\times 2$ matrix is a transfer matrix.
Values of the wave function at any lattice sites
are related to those of the initial sites 
by applying the transfer matrices successively.
There are three types of transfer matrices :
\begin{eqnarray} 
&&M_{aa}=
\left(
\begin{array}{cc}
\frac{E}{t_a}& -1\\
1 &0
\end{array}
\right), 
\nonumber\\
&&M_{ab}=
\left(
\begin{array}{cc}
\frac{E}{t_a}&-\frac{t_b}{t_a}\\
1& 0
\end{array}
\right),
\quad
M_{ba}=
\left(
\begin{array}{cc}
\frac{E}{t_b} & -\frac{t_a}{t_b}\\
1& 0
\end{array}
\right).
\end{eqnarray}
We introduce a matrix $M_{\ell}$ which generates
a wave function at the $F_{\ell}$-th site 
\begin{eqnarray}
\left(
\begin{array}{c}
\psi_{F_{\ell}+1}\\
\psi_{F_{\ell}}
\end{array}
\right)
=
M_{\ell}
\left(
\begin{array}{c}
\psi_1\\
\psi_0
\end{array}
\right).
\end{eqnarray}
This matrix $M_{\ell}$ is given by a successive product
of the transfer matrices, and
the number of element of the product is $F_{\ell}$ 
which grows like $\sigma^{-\ell}$.
A special property of the transfer matrices 
of the Fibonacci model is that  
$M_{\ell}$ is determined by a recursive map 
\cite{KohKadTan}
\begin{eqnarray}
\label{RGmap}
M_{\ell+1}=M_{\ell-1}M_{\ell},
\end{eqnarray}
with an initial condition 
$M_1=M_{aa},M_2=M_{ab}M_{ba}$.
This map simplifies a procedure to obtain
the matrix $M_{\ell}$.

\begin{figure}
  \begin{center}
    \epsfxsize=8cm
    \epsfbox{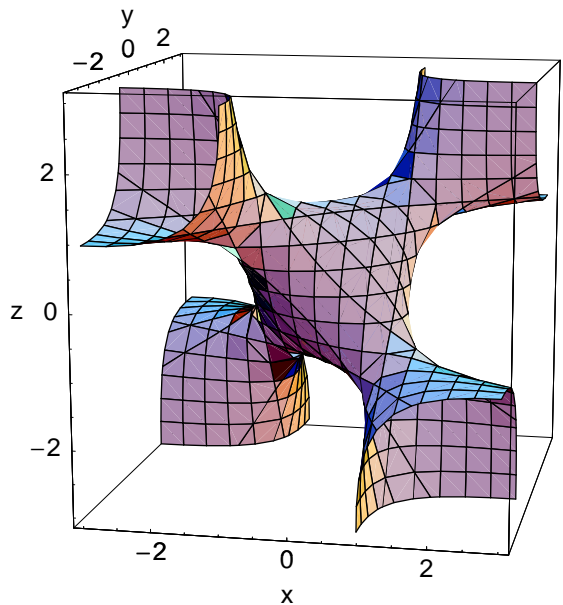}
    \caption{Two-dimensional manifold $x^2+y^2+z^2-2xyz-1=I$ 
for $I=\frac{9}{16}$.
The manifold consists of a central body and four parts to infinity.
The four parts are connected to the central body through four tubes.}
  \end{center}
\label{Manifold}
\end{figure}

As $M_{\ell}$'s are unimodular matrices, 
using the trace of the matrix $M_{\ell}$, 
the equation (\ref{RGmap}) is casted to 
\begin{eqnarray}
x_{\ell+1}=y_{\ell},\quad
y_{\ell+1}=z_{\ell},\quad
z_{\ell+1}=2y_{\ell}z_{\ell}-x_{\ell},
\label{dynamicalmap}
\end{eqnarray}
where $x_{\ell}=\frac{1}{2}{\rm Tr}M_{\ell}$ and 
we introduce three-dimensional vectors
$\vec{r}_{\ell}=(x_{\ell},y_{\ell},z_{\ell})
=(x_{\ell},x_{\ell-1},x_{\ell-2})$
\cite{KohKadTan}. This map $\vec{r}_{\ell+1} =f(\vec{r}_{\ell})$ 
is the KKT map.
An initial condition is  
$x_1=\frac{E}{2t_a},
y_1=\frac{E}{2t_b},
z_{1}=\frac{1}{2}\left(\frac{t_b}{t_a}+\frac{t_a}{t_b}\right)
$.
It is remarkable that this map has a constant of motion \cite{KohBan}:
\begin{eqnarray}
I=x_{\ell}^2+y_{\ell}^2+z_{\ell}^2-2x_{\ell}y_{\ell}z_{\ell}-1.
\end{eqnarray}
The initial condition gives the value of this constant of motion as
\begin{eqnarray}
I=\frac{1}{4}\left(\frac{t_b}{t_a}-\frac{t_a}{t_b}\right)^2.
\end{eqnarray}
The constant of motion $I$ determines a two-dimensional
manifold on which an orbit of the KKT map remains.
In Fig. 1, an example of the manifold is shown.
It is always non-compact  as long as $I>0$.
When $I$ approaches  zero, the tubes shrink to points.
A case $I < 0$ does not represent the electronic systems.

In Fig. 2, we show the energy bands for small $\ell$'s.
\begin{figure}
  \begin{center}
    \epsfxsize=8cm
    \epsfbox{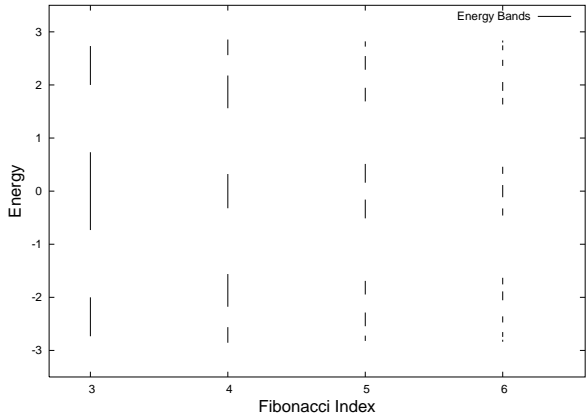}
    \caption{Energy bands at $(t_a,t_b)=(1,2)$, $I = \frac{9}{16}$
with Fibonacci numbers $F_{3}=3, F_{4}=5, F_{5}=8$ and $F_{6}=13$.
The bands consist of $F_{\ell}$ bands and $F_{\ell}-1$ gaps.
}
  \end{center}
\label{BandSplit}
\end{figure}
Since the eigenstates of the model must be normalizable, 
we impose the condition on  the initial condition so that 
$|\psi_{F_{\ell}}|$ does not grow exponentially.
In other words, the energy levels of the equation (\ref{off-diag})
with $F_{\ell}$
is determined by the condition $|x_{\ell}|\leq 1$
since the energy enters in the initial condition of the KKT map.  
It can also be shown that a condition 
$|x_i|, |x_{i+1}| >1 $ for an $i <\ell$ 
 is sufficient for the corresponding energy to be  in an energy gap. 
For large $F_{\ell}$, the KKT map provides escaping orbits on the manifold 
for almost all energies and the remaining set is shrinking 
to zero measure as $F_{\ell} \rightarrow \infty$.

\section{Level Statistics}
\hspace{2.5mm}
We use the KKT map described in the previous section 
 to get the spectrum of the Fibonacci model. 
We first consider  periodic systems 
whose unit cell is a Fibonacci number $F_{\ell}$,
then extrapolate results to understand the Fibonacci model $\ell =\infty$. 

Let us begin with the band-width distribution $P_B(w)$.
We normalize the distribution by
$\int_0^{\infty}P_B(w)dw=1$ and $\int_0^{\infty}wP_B(w)dw=1$.
Each band width  
approaches zero at limit $\ell \rightarrow \infty$ \cite{KohSutTan,KohOon}. 
At a finite  Fibonacci index $\ell$, 
we normalize the band widths by the mean of them and 
consider  distribution $P^{(\ell)}_B(w)$ of the normalized band widths. 
 Then we extrapolate  $P_B(w)$  at $\ell \rightarrow \infty$. 

Fig. 3 shows  $\ln{w}$-$\ln{P_B(w)}$ plots. 
As $ w \rightarrow 0$, 
the distribution $P_B(w)$ converges to zero.
\begin{figure}
  \begin{center}
    \epsfxsize=8cm
    \epsfbox{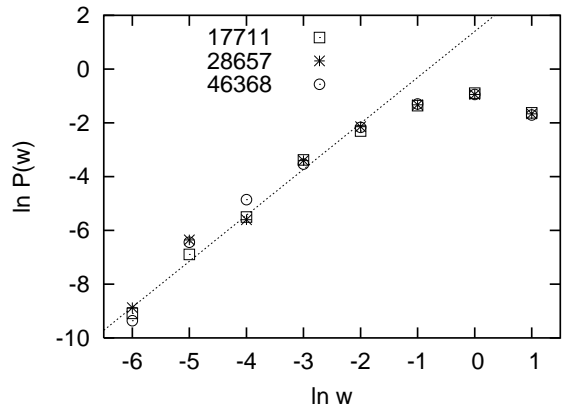}
    \caption{Band-width distributions 
at $(t_a,t_b)=(1,2)$, $I = \frac{9}{16}$.
The distributions for $\ln{w}<1.5$ 
suggest $P_B(w)\sim w^{1.71}$.}
  \end{center}
\label{loglog}
\end{figure}
The results for different $\ell$ seems to be on the same line,  
which suggests the existence of band-width 
distribution at $\ell\to\infty$.
The distribution $P_B(w)$  takes a form
\begin{eqnarray}
P_B(w)\sim w^{\alpha} \qquad (w\to 0),
\end{eqnarray}
with a positive real number $\alpha$.
This form remains to hold when $I$ changes.
The values of $\alpha$ for several cases are given in Table I.

We next consider a region where $w$ is larger than $0.5$. 
In Fig. 4, we present  distribution in $w$-$\ln{P_B(w)}$ plots. 
\begin{figure}
  \begin{center}
    \epsfxsize=8cm
    \epsfbox{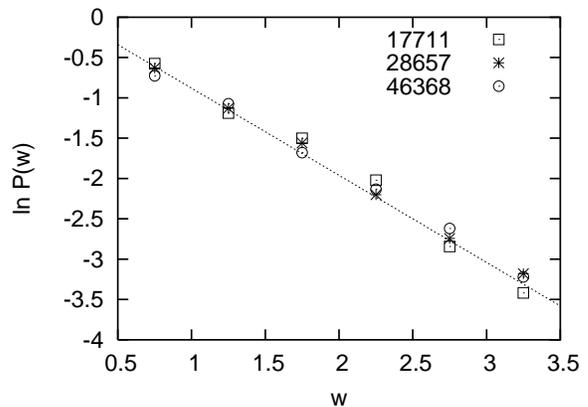}
    \caption{Band-width distributions with $w$-$\ln{P_B(w)}$ plots
at $(t_a,t_b)=(1,2)$, $I = \frac{9}{16}$.
The distributions for $w>0.5$ suggest $P_B(w)\sim e^{-1.08w}$.}   
  \end{center}
\label{semilog}
\end{figure}
The linear behavior suggests that 
the distribution $P_B(w)$ at large $w$ is given by a form
\begin{eqnarray}
P_B(w)\sim e^{-\beta w}\qquad (w\to\infty),
\end{eqnarray}
with a positive real number $\beta$.
Estimated values of $\beta$ for several cases are presented in Table I.
We have changed the hopping coefficients $t_a$ and $t_b$ for fixed $I$'s 
for several cases.
The results indicate that $\alpha$ and $\beta$ depend 
on $t_a$ and $ t_b$ only through the value of $I$.    

Next we examine the ansatz  that the distribution $P_B(w)$ is 
described by a combined distribution  
$P(w)=N_{\alpha\beta}w^{\alpha}e^{-\beta w}$ with
a normalization constant $N_{\alpha\beta}$ for every value of $w$.
This is the semi-Poisson form as
was assumed for the Harper model \cite{EvaPic}.
The normalization conditions for the band-width distribution
impose the relation between the indices $\beta=\alpha+1$.
Then the distribution reduces to generalized form of the semi-Poisson 
distribution $P(w)=\frac{(\alpha+1)^{\alpha+1}}{\Gamma(\alpha+1)}
w^{\alpha}e^{-(\alpha+1)w}$.
However this is  different from 
the obtained behavior at $w\to 0,\infty$. 
Thus we exclude the ansatz of the semi-Poisson distribution. 

\setlength{\arrayrulewidth}{0.8pt}
\begin{table}[tb]
\begin{center}
\begin{tabular}{|c|c|c|c|c|c|c|}     \hline
$t_a$ & $t_b$ &  $I$ &
 $\alpha$ & $\beta$ & $\delta$  \\  \hline

$\frac{11}{8}$  & $\frac{13}{8}$ & 
$\frac{576}{20449}$ & 
1.20($\pm$0.03)&
0.38($\pm$0.08)  &
0.935($\pm$0.004)    \\

$\frac{5}{4}$ & $\frac{7}{4}$ & 
$\frac{144}{1225}$ & 
1.28($\pm$0.05) &
1.05($\pm$0.08)  &
0.824($\pm$0.004)  \\

$\frac{9}{8}$ & $\frac{15}{8}$ &
$\frac{64}{225}$ & 
1.42($\pm$0.07) &
1.07($\pm$0.07) &
0.747($\pm$0.004)   \\

1 & 2 & 
$\frac{9}{16}$ & 
1.71($\pm$0.07) &
1.08($\pm$0.06) &
0.689($\pm$0.004)   \\

$\frac{7}{8}$ & $\frac{17}{8}$ & 
$\frac{14400}{14161}$ & 
1.87($\pm$0.13) &
1.18($\pm$0.07) &
0.612($\pm$0.002)   \\

$\frac{3}{4}$  & $\frac{9}{4}$ & 
$\frac{16}{9}$ & 
1.91($\pm$0.09)&
1.29($\pm$0.08)  &
0.555($\pm$0.003)    \\

$\frac{5}{8}$ & $\frac{19}{8}$ & 
$\frac{28224}{9025}$ & 
2.23($\pm$0.17) &
1.50($\pm$0.12)  &
0.490($\pm$0.002)  \\

\hline
\end{tabular}
\end{center}
\caption{Estimated values for critical indices $\alpha,
\beta$ and $\delta$.
Errors arise in fitting  numerical data
to the proposed distributions.}
\label{table;l2}
\end{table}

We found that the gap distribution $P_G(s)$ 
always  diverge for $s\to 0$.
An example  using the $\ln{s}$-$\ln{P_G(s)}$ plot
is shown in Fig. 5.
\begin{figure} 
  \begin{center}
    \epsfxsize=8cm
    \epsfbox{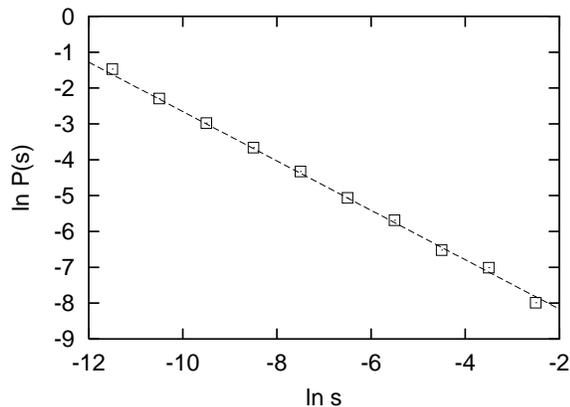}
    \caption{Gap distribution  
$\ln{s}$-$\ln{P_G(s)}$ 
for $(t_a,t_b)=(1,2)$, $I= \frac{9}{16}$ and  $F_{21}=17711$.
$P_G(s)\sim s^{-0.689}$.}
  \end{center}
\label{gap}
\end{figure}
We normalize the  distribution as $\int_0^{\infty}P_G(s)ds=1$. 
From  Fig. 5, the 
 gap distribution 
obeys an inverse power law  
\begin{eqnarray}
P_G(s)\sim s^{-\delta} \qquad (s\to 0),
\end{eqnarray}
which diverges at the origin.
Values of $\delta$ for several cases are given in Table I. 
We have changed the hopping coefficients $t_a$ and $t_b$ for fixed $I$'s 
for several cases.
The results indicate that
$\delta$ depends on $t_a$ and $ t_b$ only through $I$.

\section{Comparison with the multi-scale Cantor sets}
\hspace{2.5mm}
We compare our results in the previous section
 with properties of  multi-scale 
Cantor sets which are produced by splitting an interval 
into a definite number of 
subintervals with definite scales (see Appendix for details).   
 We call the intervals of the multi-scale Cantor sets  
 ``bands'',
while we call the removed intervals as ``gaps''.  

Let us consider the band-width distribution of a multi-scale Cantor set,
$P_B^{({\rm Cantor})}(w)$. 
As far as we know, 
$P_B^{({\rm Cantor})}(w)$ 
has not been appeared in the literature.  
However, Kolmogorov's theory of turbulence \cite{Kol} 
introduces the energy distribution measured at a definite spatial scale, 
which is conceptually equivalent to $P_B^{({\rm Cantor})}(w)$ of our case. 
We derive  
$P_B^{({\rm Cantor})}(w)$ in 
 the Appendix, and it turns out to follow  
a logarithmic normal distribution as a distribution 
in Kolmogorov's theory, 
\begin{eqnarray}
P_B^{({\rm Cantor})}(w)
\sim
\frac{1}{w}N_{(0,n\sigma^2)}(\ln{w}),
\qquad (n\to\infty),
\end{eqnarray}
where $N_{(\mu,\sigma^2)}(x)
=\frac{1}{\sqrt{2\pi}\sigma}e^{-\frac{(x-\mu)^2}{2\sigma^2}}$ 
and 
$\sigma^2$ is the variance of the logarithm of  
the scales to generate a multi-scale Cantor set and  $n$ 
is the number of steps to produce the set. 
Thus for a  multi-scale Cantor set, $P_B^{({\rm Cantor})}(w)$ 
becomes broader for large $n$ 
and does not converge to a normalizable distribution. 
This contrasts  with the behavior of $P_B(w)$ for the Fibonacci model.

As shown in the Appendix,
we can modify the construction of a multi-scale Cantor set so that
$P_B^{({\rm Cantor})}(w)$ has a constant variance 
by including a finite size correction to the scales 
at each splitting step. 
When we write the width of a band
 produced at step $k$ as $w_k=e^{-k\epsilon_k}$, 
$\epsilon_k$ would receive a correction proportional to $\frac{\ln k}{k}$.
Then
\begin{eqnarray}
P_B^{({\rm Cantor})}(w)=\frac{1}{w}N_{(0,\sigma^2)}(\ln{w}).
\end{eqnarray}
A graph for a logarithmic normal distribution is shown in Fig. 6. 
\begin{figure}
\begin{center}
    \epsfxsize=8cm
    \epsfbox{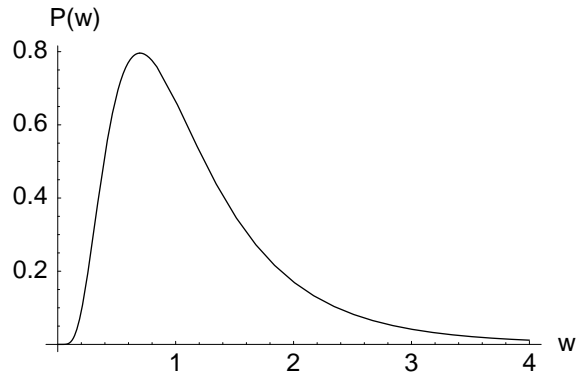}
    \caption{Logarithmic normal distribution 
$P(w)=\frac{1}{w}N_{(\mu,\sigma^2)}(\ln{w})=
\frac{1}{\sqrt{2\pi}\sigma}\frac{1}{w}
e^{-\frac{(\log{w}-\mu)^2}{2\sigma^2}}$ with $\mu=0,\sigma=0.6$}
\end{center}
\label{log-normal}
\end{figure}
At the origin, the logarithmic normal distribution approaches 
zero faster than power law. 
Also at the tail, it does not decay at an exponential rate.
Thus, the behavior of $P_B^{({\rm Cantor})}(w)$ of this multi-scale
 Cantor set with the finite size corrections
shows differences from those of the Fibonacci model. 

These differences may be rooted in the degree of self-similarity.
The multi-scale Cantor sets are genuinely 
self-similar : the same splitting procedure is 
applied to every interval at each step. This leads to a 
logarithm normal contribution.  
On the other hand, the splitting of the energy levels of
the Fibonacci model is different for each band 
and it has  different scales for different bands
\cite{KohKadTan,KohOon}. 
In \cite{KohOon}, it has been observed that 
numerical values of 
scales of the band's center coincide with 
eigenvalues  of 
the linearized KKT map at the hyperbolic points of 
  the corresponding periodic orbit and 
the splitting of the bands is explained by the Smale's 
horseshoe structure. 
This observation can be applied to other periodic orbits corresponding to 
other bands.
Then, at least, some subsets of the bands are self-similar and 
should satisfy the band-width distribution 
like the multi-scale Cantor sets.  
Still a finite number of periodic orbits 
only lead to the sum of logarithmic normal distributions and  
cannot explain fully the behavior of the Fibonacci model. 

Then our comparison implies that, to understand the behavior of 
$P_B(w)$ of the Fibonacci model by the language of dynamical 
system, it is necessary to include  
finite size corrections as well as an infinite number of 
periodic orbits. 
Such a fine structure of fractal is not taken into account  
in multifractal analysis of \cite{Halsey}, which characterizes 
fractal objects by the statistical distribution of scales. 
As proved in \cite{Suto}, the energy spectra of the Fibonacci model
belong to a class of a general Cantor set. 
Thus they belong to a different class of Cantor sets 
to the multi-scale Cantor sets.

We also compare the results for  the gap distribution.
The gap distribution of the multi-scale Cantor set
follows an inverse power law near the origin $s\to 0$
\begin{eqnarray}
P_G^{({\rm Cantor})}(s)\sim s^{-\delta}, \qquad \delta=D_H+1,
\label{gapcantor}
\end{eqnarray} 
where $D_H$ is the Hausdorff dimension. 
This is because the length of gaps of a multi-scale Cantor set is directly 
related to the length of the removed elements since 
the bands and gaps are complementary in a fixed interval. 
Although 
the Fibonacci model have an inverse power  
law like Eq.(\ref{gapcantor}) for the gap distribution, 
 the relation $\delta= D_H+1$ does not hold for the Fibonacci model 
 because of  the value of the Hausdorff dimension
$\sim 0.685$ \cite{KohSutTan}.  
This discrepancy is due to the splitting 
of the energy levels as observed in Fig. 1: 
the energy levels of the Fibonacci model
have fluctuating endpoints, 
and the bands and the gaps are not complementary of 
a fixed interval, and hence have different scaling properties.  
On the other hand, for the Harper model,   
the numerical results of  \cite{TakInoYam,MachiGei} seem to 
sustain the relation $\delta=D_H+1$. 
From the same reasoning, 
there seems to exist no explanation for it actually.
 
\section{Conclusion}
\hspace{2.5mm}
Spectral properties of the Fibonacci model
are examined numerically by using the Kohmoto-Kadanoff-Tang map. 
We investigated the band-width distribution
$P_B(w)$ and the gap distribution $P_G(s)$.
We find a power law $P_B(w) \sim w^{\alpha}$ near the origin and 
an exponential decay $P_B(w) \sim e^{-\beta w}$ 
toward  infinity ($\alpha,\beta>0$). 
Also $P_G(s)$ diverges  
with an inverse power law $s^{-\delta}$ near the origin ($\delta>0$).
The indices turn out to depend on the coupling constants
through the constant of motion in the Kohmoto-Kadanoff-Tang map.

We compare the band-width distribution $P_B(w)$ and 
the gap distribution $P_G(s)$ of the Fibonacci model
 with those of  multi-scale Cantor sets. 
As the distribution of the ``band" widths  
$P_B^{({\rm Cantor})}(w)$ for  multi-scale Cantor sets 
always obeys a logarithmic normal distribution, 
the behavior of $P_B(w)$ is rather different 
from that of $P_B^{({\rm Cantor})}(w)$. 
Also, while the distribution of the ``gaps''
$P_G^{({\rm Cantor})}(s)$ of 
the multi-scale Cantor sets 
is an inverse power law $s^{-\delta}$ near the origin ($\delta>0$)
and there is the relation between the index $\delta$ 
and its Hausdorff dimension,  
the relation does not hold for the Fibonacci model. 
Thus the critical level statistics is useful 
to quantify differences between the energy spectra of 
the Fibonacci models and  multi-scale Cantor sets 
which are hidden  in the language of multifractal analysis.

\vspace{2ex}

{ \bf Acknowledgments} 
\hspace{2.5mm}
M.N. would like to thank Y. Takada for discussion.
The research of M.N. is supported by JSPS fellowship (No.0206911).
K.I. has benefited from 
the Grand-in-Aid for Science(B), No.1430114  of JSPS.

\appendix
\section*{Appendix: Distributions for the multi-scale Cantor sets}
\hspace{2.5mm}
The general definition of a Cantor set is a closed set with 
no isolated points and whose complement is dense. 
Here we restrict ourselves 
 to the so-called multiple-scale Cantor sets. 
A multi-scale Cantor set is  a set which is obtained 
by the following procedure : we transform the interval $I=[0,1]$ into 
a finite number of subintervals with definite scales 
and apply the same transformation to subintervals, 
and the limit set is called a multi-scale Cantor set.

A standard example is the classical Cantor set : 
We split an interval into three equal pieces 
and remove the middle one. 
Two remaining intervals are $[0,\frac{1}{3}]$ and $[\frac{2}{3},1]$. 
One can continue the same transformation for these intervals. 
The number of intervals which we call "band" at step $n$ is given 
by $N=2^{n}$ with  length $W=\frac{1}{3^n}$.
The removed intervals, which we call  ``gap'', have 
lengths $S=\frac{1}{3},\frac{1}{3^2},\cdots,\frac{1}{3^n}$.   
As $n$ goes to $\infty$, we end up with 
an infinite number of infinitesimal intervals. 
The Hausdorff dimension of the  set is given 
by $D_H=\frac{\ln N}{\ln\frac{1}{W}}=\frac{\ln 2}{\ln 3}$.

For a multi-scale Cantor set, 
let $n$ be the number of steps  to generate the set 
and $N$ be the number of bands : 
$N \sim a^{n}$ with a number of split intervals $a$.    
Let $W_i,S_i$ $(i=1,\cdots, N)$
be the length of bands and gaps respectively.
We normalize $W_i$  so that their mean is equal to 1.  
The mean of $W_i$ at step $n$ is given by 
$\overline{W}_n = \frac{1}{N}\sum_{i=1}^{N} W_i$. 
The normalized band-widths are  
$ w_i := \frac{W_i}{\overline{W}_n}$.

For the Cantor set with one scale described above, 
$N=2^n, W_i=3^{-n},N\overline{W}=\sum_{i=1}^{N}W_i = \frac{2^n}{3^n},
 \overline{W} = \frac{1}{3^n}$ and $w_i = 1$, 
we have  
\begin{eqnarray}
P_B^{({\rm Cantor})}(w) =\delta(w-1).
\end{eqnarray}
This means that there is no distribution of scaling variables and 
is readily generalized to a scale other than $\frac{1}{3}$.  
On the other hand, the gap lengths are  
$S_i=\frac{1}{3},\frac{1}{3^2},\cdots,\frac{1}{3^n}$. 
The number of the gaps of $\frac{1}{3^k}$ ($k=1,\cdots,n$) is $2^k$. 
Thus the distribution of the gaps is given by
\begin{eqnarray}
P_G^{({\rm Cantor})}(s)\sim s^{-\frac{\ln 2}{\ln 3}-1}=s^{-D_H-1}.
\end{eqnarray}  
The power is related to the Hausdorff dimension
because the lengths of the gaps 
consist of those of the band-widths.

Let us next consider a two-scale Cantor set which has 
 two scales
$r_1,r_2>0$ with $r_1+r_2=\rho_0<1$. 
Namely we split an interval into three intervals 
with length $r_1,r_2$ and $t=1-\rho_0$, 
and remove the interval with the length $t$.
Then we continue the same procedure to the remaining intervals.  
We put $N=2^n$, $\rho=\rho_0/2$ and 
define $\widetilde{x}=x/\rho, \widetilde{y}=y/\rho$. 

For the bands, we have 
\begin{eqnarray}
W_i =r_1^n,r_1^{n-1}r_2,r_1^{n-2}r_2^2,\cdots,r_1r_2^{n-1},r_2^n,
\end{eqnarray}
where each of them has ${n \choose m}$ terms. 
Then, the mean of the band-widths is given by
\begin{eqnarray} 
N\Delta &=& \sum_{m} {n \choose m}r_1^{n-m}r_2^m 
= (r_1+r_2)^n = \rho_0^n, \nonumber\\ 
\Delta &=&  \rho^n.
\end{eqnarray}
The normalized band-widths are
\begin{eqnarray}
w_i &=& 
\widetilde{r}_1^n,\widetilde{r}_1^{n-1}\widetilde{r}_2,
\widetilde{r}_1^{n-2}\widetilde{r}_2^2,\cdots,
\widetilde{r}_1\widetilde{r}_2^{n-1},\widetilde{r}_2^n. 
\end{eqnarray}
Since the binomial distribution ${n \choose x}$ 
converges to a normal distribution as $n \rightarrow \infty$, 
when $n$ is taken as a discrete time, $x$ can be regarded as 
a discrete Brownian motion (a binomial model of it).
To get $P_B^{({\rm Cantor})}(w)$ for $n \rightarrow \infty$, 
let us recall the central limit theorem: 

[Theorem] (The Central Limit Theorem) 

{\it Let $X_k$ be  independent stochastic variables following 
an identical probabilistic distribution. 
Let $E(X_1)=\mu, Var(X_1)=\sigma^2$. Then 
\begin{eqnarray}
&&\lim_{n \rightarrow \infty}{\rm Prob} 
\left[ \frac{X_1+X_2+\cdots+X_n-n\mu}{\sigma\sqrt{n}} \leq x \right]
= \Phi(x), \nonumber \\
&&\Phi(x) = \int_{-\infty}^{x}N_{0,1}(y)dy, 
\; N_{(\mu, \sigma^2)}(y)=
\frac{1}{\sqrt{2\pi}\sigma}e^{-\frac{(y-\mu)^2}{2\sigma^2}}.
\nonumber\\
&&~~~~~~~~~~~~~~~~~~~~~~~~~~~~~~~~~~~~~~~~~~~~~~~~~~~~~~
~~~~~~~~~~\Box
\nonumber 
\end{eqnarray}}
In our case,   we define $X_k=\ln \frac{W_k}{W_{k-1}}$ ($W_0=1$). 
$X_k$ take values $u=\ln r_1, d=\ln r_2$, and
$E(X_k)=\frac{u+d}{2}, Var(X_k) = \frac{(u-d)^2}{4}$.
Then the central limit theorem applies : 
\begin{equation}
\lim_{n \rightarrow \infty} {\rm Prob}
\left[ \frac{\ln \frac{W_n}{W_0} -\frac{n(u+d)}{2}}{\frac{1}{2}(u-d)
\sqrt{n}} \leq x \right] = \Phi(x).
\end{equation}
This means that $\ln \frac{W_n}{W_0}$  obeys the distribution 
$N_{(\frac{n}{2}(u+d),\frac{n(u-d)^2}{4})}$.
Returning to the normalized variables $w_i$, we similarly define 
$x_k = \ln \frac{w_k}{w_{k-1}}$.
The band-width distribution is obtained as 
\begin{equation} 
P_B^{({\rm Cantor})}(w) 
\sim
\frac{1}{w}N_{(0,\frac{n(u-d)^2}{4})}(\ln w),
\qquad (n\to\infty).
\label{pbwcantor}
\end{equation}
This is a logarithmic normal distribution
whose variance increases linearly with $n$.  

The distribution (\ref{pbwcantor}) can be generalized 
to Cantor sets with arbitrary number of  scales 
as long as the mean and the variance of the scales are finite. 
In this case,  
the central limit theorem applies.  
$P_B^{({\rm Cantor})}(w)$ always 
obeys a logarithmic normal distribution with a variance 
proportional to $n$ :
\begin{eqnarray}
P_B^{({\rm Cantor})}(w)
\sim
\frac{1}{w}N_{(0,n\sigma^2)}(\ln{w}),
\qquad (n\to\infty),
\end{eqnarray}
where $\sigma^2$ is the variance of the logarithm of the scales.

To obtain $P_B^{({\rm Cantor})}(w)$ with a definite variance, 
one can modify the transformation of splitting a interval. 
For simplicity, we describe it for the two-scale Cantor set. 
We divide the interval $[0,1]$ into three intervals 
with lengths $r_1,r_2$ and $1-\rho_0$ with 
$r_1,r_2>0$, $r_1+r_2=\rho_0<1$, and 
remove the interval with the length $1-\rho_0$.
We continue the procedure for the remaining intervals 
but, at  step $k$, we always replace $r_1$ and $r_2$ with 
$r_1(k)=r_1^{\frac{1}{\sqrt{k}}-\frac{1}{\sqrt{k-1}}}$
and $r_2(k)=r_2^{\frac{1}{\sqrt{k}}-\frac{1}{\sqrt{k-1}}}$. 
When we write the width of a band produced 
at step $k$ as $w_k=e^{-k\epsilon_k}$
by an exponent $\epsilon_k$, the factor is amount to change $\epsilon_k$
by a finite size correction $\frac{1}{2}\frac{\ln{k}}{k}$. 
Then the distribution becomes a logarithmic normal distribution with 
a constant variance:
\begin{equation} 
P_B^{({\rm Cantor})}(w) =\frac{1}{w}N_{(0,\frac{(u-d)^2}{4})}(\ln w). 
\end{equation}
This can be readily generalized to the cases with more than two scales, 
leading to the distribution 
\begin{eqnarray}
P_B^{({\rm Cantor})}(w)=\frac{1}{w}N_{(0,\sigma^2)}(\ln{w}), 
\end{eqnarray}
where $\sigma^2$ is the variance of the logarithm of the scales.

For the gap distribution of the multi-scale Cantor sets,  
the formula $\int^{s}_0 P_G^{({\rm Cantor})}
(s') ds'\sim s^{-D_H}$ 
with their Hausdorff dimension $D_H$ holds generally
because the lengths of the gaps are directly connected 
with those of the band-widths.
Thus the gap distribution for the multi-scale Cantor sets is given by
\begin{eqnarray}
P_G^{({\rm Cantor})}(s) \sim s^{-D_H-1}.
\end{eqnarray} 



\begin{thebibliography}{99}

\bibitem{LeeRam}
P. A. Lee and T. V. Ramakrishnan,
Rev. Mod. Phys. {\bf 57}, 287 (1985). 

\bibitem{HirKoh} For a review, see
H. Hiramoto and M. Kohmoto,
Int. J. Mod. Phys. {\bf B6}, 281 (1992).


\bibitem{AubAnd}  
S. Aubry and G. Andr\'{e},
Ann. Israel Phys. Soc. {\bf 3}, 133 (1980).


\bibitem{Hof} 
D.R. Hofstadter, 
Phys. Rev.{\bf  B14}, 2239 (1976).

\bibitem{Thouless} 
D.J. Thouless, 
Phys. Rev.{\bf B28}, 4272 (1983);
Commun. Math. Phys. {\bf 127}, 187 (1990).

\bibitem{Koh}
M. Kohmoto, Phys. Rev. Lett. {\bf 51}, 1198 (1983).

\bibitem{Halsey}
T.C. Halsey, M.H. Jensen, L.P. Kadanoff, I. Procaccia and B.I. Shraiman, 
Phys. Rev. {\bf A33}, 1141 (1986); 
M. Kohmoto, Phys. Rev. {\bf A37}, 1345 (1988).

\bibitem{EvaPic}
N. Evangelou and J.-L. Pichard,
Phys. Rev. Lett. {\bf 84}, 1643 (2000).

\bibitem{TakInoYam}
Y. Takada, K. Ino and M. Yamanaka,
Phys. Rev. {\bf E} in press,  [cond-mat/0312650].

\bibitem{MachiGei}  
K. Machida and M. Fujita, 
Phys. Rev. {\bf B34}, 7367 (1986); 
T. Geisel, R. Ketzmerick and G. Petschel,
Phys. Rev. Lett. {\bf 66}, 1651 (1991).

\bibitem{KohKadTan}
M. Kohmoto, L.P. Kadanoff and C. Tang,
Phys. Rev. Lett. {\bf 50}, 1870 (1983).

\bibitem{Ostlund} 
S. Ostlund, R. Pandits, D. Rand, H.J. Schellnhuber and E. Siggia, 
Phys. Rev. Lett. {\bf 50}, 1873 (1983).

\bibitem{KohBan}
M. Kohmoto and J.R. Banavar,
Phys. Rev. {\bf B34}, 563 (1986).

\bibitem{KohSutTan}
M. Kohmoto, B. Sutherland and C. Tang,
Phys. Rev. {\bf B35}, 1020 (1987).

\bibitem{KohOon}
M. Kohmoto and Y. Oono,
Phys. Lett. {\bf 102A}, 145 (1984).

\bibitem{Suto}
A. S\"ut$\tilde{{\rm o}}$,
J. Stat. Phys. ${\bf 56}$, 525 (1989).

\bibitem{Kol} 
A.N. Kolmogorov, J. Fluid. Mech. {\bf 13}, 82 (1962).





\end{thebibliography}
\end{document}